\documentstyle[amssymb,preprint, aps]{revtex}

\begin{document}
\title{Quasi-spin Model for Macroscopic Quantum Tunnelling between Two Coupled
Bose-Einstein Condensates}
\author{Chaohong Lee$^{\thanks{%
Corresponding author: chlee@mpipks-dresden.mpg.de and chleecn@hotmail.com.}}$}
\address{Wuhan Institute of Physics and Mathematics, The Chinese Academy of Sciences,
Wuhan 430071, P. R. China, and \\
Max Planck Institute for the Physics of Complex Systems, Noethnitzer Str.
38, D-01187 Dresden, Germany}
\author{Wenhua Hai}
\address{Department of Physics, Hunan Normal University, Changsha 410081,P. R. China}
\author{Xueli Luo, Lei Shi, and Kelin Gao}
\address{Wuhan Institute of Physics and Mathematics, The Chinese Academy of Sciences,
Wuhan 430071, P. R. China}
\date{\today}
\maketitle

\begin{abstract}
The system of two coupled Bose-Einstein condensates is mapped onto a
uniaxial spin with an applied magnetic field. The mean-field interaction,
the coupling and the asymmetry or the detuning correspond to the anisotropy,
the transverse field, and the longitudinal field, respectively. A
generalized Bloch equation is derived. In the low barrier limit for the
quasi-spin model, the tunneling rate is analyzed with an imaginary-time
path-integral method. The dependence of the tunneling rate on the system
parameters is obtained. The crossover temperature $T_{c}$ from the thermal
regime to the quantum regime is estimated. Below $T_{c}$ quantum tunnelling
prevails, otherwise thermal activation dominates.
\end{abstract}



\section{Introduction}

The experimental realization of measuring the relative phase and the
population oscillation between coupled Bose-Einstein condensates (BECs)
stimulates great interest in investigating their macroscopic quantum
tunnelling dynamics \cite{Anglin,Leggett,Book}. There are two different
types of atomic tunnelling between coupled BECs, external tunnelling and
internal tunnelling \cite{Leggett,Book}. The former has different spatially
separated single-particle states in a double-well or multi-well potential
and the latter has different hyperfine internal states in a single-well
potential. For external tunnelling, the phase interference between BECs
confined in a multi-well potential has been observed \cite{Orzel,Andrews};
the experimental observation of the tunnelling among BECs confined in
multi-well potential has also been reported \cite
{Greiner,Cataliotti,Anderson}. For internal tunnelling, JILA realized a
two-component BEC coupled with Raman pulses \cite{Hall}, MIT observed the
tunnelling across spin domains in BECs \cite{Stamper-Kurn,Stenger}, and LENS
reported the current-phase dynamics in two weakly coupled BECs trapped in
different Zeeman states \cite{INFM}.

With the proceeding of the experimental exploration, a lot of theoretical
investigation was performed simultaneously. Williams et al. demonstrated the
existence of Josephson tunnelling in a driven two-state single-particle BEC
in a single-well trap potential \cite{Williams}. Kasamatsu et al.
investigated theoretically the existence of a metastable state and the
possibility of decay to the ground state through macroscopic quantum
tunnelling in two-component BECs with repulsive interactions \cite{Kasamatsu}%
. Smerzi et al. studied the coherent atomic tunneling and population
oscillations between two zero-temperature BEC's confined in a double-well
potential \cite{Smerzi1,Raghavan,Marino,Smerzi2}. Macroscopic quantum
self-trapping (MQST), namely a self-maintained population imbalance with
nonzero average value of the fractional population imbalance, and $\pi -$%
phase oscillations in which the time averaged value of the phase difference
is equal to $\pi $ were detailed in Refs. \cite{Smerzi1,Raghavan}. The
authors of Ref.\cite{Marino} claim that interaction with a thermal cloud
will damp all different oscillations to the zero-phase mode. In addition,
macroscopic quantum fluctuations have also been discussed by using
second-quantization approaches \cite{Smerzi2,Kuang}. Within the
time-dependent potential, chaotic population tunnelling emerges. Abdullaev
and Kraenkel analyzed the nonlinear resonances and chaotic oscillations of
the fractional population imbalance between two coupled BEC's in a
double-well trap with a time-dependent tunneling amplitude for different
damping \cite{Abdullaev1}. They also considered the chaotic atomic
population resonances and the possibility of stabilization of the
unstable-mode regime in coupled BEC's with oscillating atomic scattering
length \cite{Abdullaev2}. In a previous paper, we investigated the chaotic
and frequency-locked population oscillation between two coupled BECs \cite
{Lee}.

Although many papers appear in the field of the tunnelling between coupled
BECs, because of the nonlinearity in the Gross-Pitaevskii equation (GPE),
few of them address the question of calculating the tunnelling rate and the
crossover temperature between different tunnelling regimes. However, the
tunnelling rate and the crossover temperature of the spin systems have been
studied systematically with the imaginary-time path-integral method,
including models with applied magnetic field \cite
{Chudnovsky1,Chudnovsky2,Chudnovsky3,Zaslavskii,Scharf,Garanin,Garanin1} and
without \cite{Liang,Delft,Loss}. For a two-state system described with
linear Schr$\stackrel{..}{o}$dinger equation, it is easy to visualize the
effects of coupling between two states by introducing Bloch's spin vector
formalism \cite{Allen}. Can we introduce a generalized Bloch vector for two
coupled BECs described with the nonlinear Schr$\stackrel{..}{o}$dinger
equation to map it onto a spin system, and then calculate the tunneling rate
and the crossover temperature with the imaginary-time path-integral method?
If the coupled BECs is equivalent to a spin system, the tunnelling process
is related to the decay of the metastable MQST state to the ground state.
More interestingly, the crossover temperature corresponds to the transition
from the classical or mean-field regime to the second quantization regime.
In the next section, by introducing a generalized Bloch spin vector, the
coupled BECs is mapped onto an uniaxial spin with an applied magnetic field.
In section III, the tunneling rate is calculated with the imaginary-time
path-integral method, and the crossover temperature is estimated. In the
last section, a brief discussion and summary is given.

\section{Quasi-spin model for two coupled Bose-Einstein condensates}

Consider the experiments of JILA \cite{Hall}, two Bose-Einstein condensates
in the $|F=1,m_{F}=-1>=|1>$ and $|F=2,m_{F}=1>=|2>$ spin states of $%
^{87}R_{b}$ are coupled by a two-photon pulse with the two-photon
Rabi-frequency $\Omega $ and a finite detuning $\delta =\omega _{d}-\omega
_{hf}$. Where, $\omega _{d}=\omega _{1}+\omega _{2}$ is the driven frequency
of the two-photon pulses, $\omega _{hf}$ is the transition frequency between
two hyperfine states. In the rotating frame, ignoring the damping and the
finite-temperature effects, the coupled two-component BEC system can be
described by a pair of coupled GPEs 
\begin{equation}
\begin{array}{l}
i\hbar \frac{\partial \Psi _{2}(\stackrel{\rightharpoonup }{r},t)}{\partial t%
}=(H_{2}^{0}+H_{2}^{MF}-\frac{\hbar \delta }{2})\Psi _{2}(\stackrel{%
\rightharpoonup }{r},t)+\frac{\hbar \Omega }{2}\Psi _{1}(\stackrel{%
\rightharpoonup }{r},t), \\ 
i\hbar \frac{\partial \Psi _{1}(\stackrel{\rightharpoonup }{r},t)}{\partial t%
}=(H_{1}^{0}+H_{1}^{MF}+\frac{\hbar \delta }{2})\Psi _{1}(\stackrel{%
\rightharpoonup }{r},t)+\frac{\hbar \Omega }{2}\Psi _{2}(\stackrel{%
\rightharpoonup }{r},t),
\end{array}
\end{equation}
where, the free evolution Hamiltonians $H_{i}^{0}=-\frac{\hbar
^{2}\triangledown ^{2}}{2m}+V_{i}(\stackrel{\rightharpoonup }{r})$ $(i=1,2)$
and the mean-field interaction Hamiltonians $H_{i}^{MF}=\frac{4\pi \hbar ^{2}%
}{m}(a_{ii}|\Psi _{i}(\stackrel{\rightharpoonup }{r},t)|^{2}+a_{ij}|\Psi
_{j}(\stackrel{\rightharpoonup }{r},t)|^{2})$ $(i,j=1,2$, $i\neq j)$. The
coefficient $a_{ij}$ is the scattering length between states $i$ and $j$ and
it satisfies $a_{ij}=a_{ji}$. Weak coupling is defined by the Rabi frequency
satisfying $\Omega /(\omega _{x}\omega _{y}\omega _{z})^{1/3}=\Omega /%
\overline{\omega }\ll 1$, where $\overline{\omega }=(\omega _{x}\omega
_{y}\omega _{z})^{1/3}$ is the geometric-averaged angular frequency for the
trapping potential. In this regime, we can write the macroscopic
wavefunctions using the variational ansatz $\Psi _{i}(\stackrel{%
\rightharpoonup }{r},t)=\psi _{i}(t)\Phi _{i}(\stackrel{\rightharpoonup }{r})
$ with $\psi _{i}(t)=\sqrt{N_{i}(t)}e^{i\alpha _{i}(t)}$ $(i=1,2)$. In the
ansatz, the functions $\Phi _{i}(\stackrel{\rightharpoonup }{r})$ describe
the spatial distribution of the $i-th$ component, the complex coefficient
functions $\psi _{i}(t)$ are spatially uniform and contain all
time-dependence in the macroscopic quantum wave-functions $\Psi _{i}(%
\stackrel{\rightharpoonup }{r},t)$. The symbols $N_{i}(t)$ and $\alpha
_{i}(t)$ represent the populations and phases of the $i-th$ condensate,
respectively. Because the coupling is very weak, the spatial distributions
vary slowly in time and are very close to the adiabatic solutions to the
time-independent uncoupled case for GP equations $(1)$, being slaved by the
populations\cite{Williams}. Thus, the complex coefficient functions $\psi
_{i}(t)$ obey the nonlinear two-mode dynamical equations 
\begin{equation}
\begin{array}{l}
i\hbar \frac{d}{dt}\psi _{_{2}}(t)=(E_{2}^{0}-\frac{\hbar \delta }{2}%
+U_{22}|\psi _{_{2}}(t)|^{2}+U_{21}|\psi _{_{1}}(t)|^{2})\psi _{_{2}}(t)+%
\frac{K}{2}\psi _{_{1}}(t), \\ 
i\hbar \frac{d}{dt}\psi _{_{1}}(t)=(E_{1}^{0}+\frac{\hbar \delta }{2}%
+U_{11}|\psi _{_{1}}(t)|^{2}+U_{12}|\psi _{_{2}}(t)|^{2})\psi _{_{1}}(t)+%
\frac{K}{2}\psi _{_{2}}(t).
\end{array}
\end{equation}
The parameters satisfy $E_{i}^{0}=\int \Phi _{i}(\stackrel{\rightharpoonup }{%
r})H_{i}^{0}\Phi _{i}(\stackrel{\rightharpoonup }{r})d\stackrel{%
\rightharpoonup }{r}$, $U_{ij}=\frac{4\pi \hbar ^{2}a_{ij}}{m}\int |\Phi
_{i}(\stackrel{\rightharpoonup }{r})|^{2}|\Phi _{j}(\stackrel{%
\rightharpoonup }{r})|^{2}d\stackrel{\rightharpoonup }{r}=U_{ji}$ and $%
K=\hbar \Omega \int \Phi _{1}(\stackrel{\rightharpoonup }{r})\Phi _{2}(%
\stackrel{\rightharpoonup }{r})d\stackrel{\rightharpoonup }{r}$ $(i,j=1,2)$.
The terms in $K$ describe population transfer (internal tunnelling) between
two BEC states, whereas the terms in $U_{ij}$, which depend on the numbers
of atoms in each BEC state, describe the mean-field interaction between
atoms. When $U_{21}$ and $\delta $ equal zero, these coupled equations can
also describe the BECs in a double-well potential \cite
{Smerzi1,Raghavan,Marino,Smerzi2}. Similar to the coupled two-state system
obeying the linear Schr$\stackrel{..}{o}$dinger equation, we introduce a
generalized Bloch spin vector $(u,v,w)$ with the components 
\begin{equation}
u=\psi _{_{2}}^{*}\psi _{_{1}}+\psi _{_{2}}\psi _{_{1}}^{*}\text{ },\text{ }%
v=-i(\psi _{_{2}}\psi _{_{1}}^{*}-\psi _{_{2}}^{*}\psi _{_{1}})\text{ },%
\text{ }w=\psi _{_{2}}^{*}\psi _{_{2}}-\psi _{_{1}}^{*}\psi _{_{1}}.
\end{equation}
Obviously, $u^{2}+v^{2}+w^{2}=(N_{1}+N_{2})^{2}=N_{T}^{2}$ is a conserved
quantity when finite-temperature and damping effects can be ignored.
Rescaling the time $t/\hbar $ to $t$, the Bloch spin vector satisfies 
\begin{equation}
\frac{du}{dt}=v(\gamma +\eta w)\text{ },\text{ }\frac{dv}{dt}=Kw-u(\gamma
+\eta w)\text{ },\text{ }\frac{dw}{dt}=-Kv,
\end{equation}
where $\gamma =E_{2}^{0}-E_{1}^{0}+N_{T}(U_{22}-U_{11})/2-\hbar \delta $ and 
$\eta =(U_{22}+U_{11}-2U_{12})/2$. Regarding the atom in one condensate as
spin-up state and the atom in the other condensate as spin-down state, the
coupled BECs can be described with the quasi-spin $\stackrel{\rightharpoonup 
}{S}=u\stackrel{\rightharpoonup }{e_{x}}+v\stackrel{\rightharpoonup }{e_{y}}%
+w\stackrel{\rightharpoonup }{e_{z}}$. In this language, the longitudinal
component $w$ depicts the population difference, and the transverse
components $u$ and $v$ characterize the coherence. Thus the effective
Hamiltonian for the quasi-spin is 
\begin{equation}
E=-\frac{1}{2}\eta S_{z}^{2}-KS_{x}-\gamma S_{z}.
\end{equation}
The above Hamiltonian is similar to the one of a uniaxial spin with an
applied magnetic field \cite{Chudnovsky3,Zaslavskii,Scharf,Garanin,Garanin1}%
, it indicates that the mean-field interaction brings the anisotropy $\eta $%
, the coupling causes an effective transverse magnetic field $K$ along axis-$%
x$, and the asymmetry or the detuning induces an effective longitudinal
magnetic field $\gamma $. In the symmetric case $(E_{2}^{0}=E_{1}^{0},$ $%
U_{22}=U_{11}$ and $\delta =0)$, it is consistent with the one derived from
the second quantized Hamiltonian in \cite{Orzel}.

\section{Tunnelling rate and crossover temperature}

In conventional spherical coordinates, the spin components can be written as 
$S_{x}=N_{T}\sin \theta \cos \phi ,S_{y}=N_{T}\sin \theta \sin \phi $ and $%
S_{z}=N_{T}\cos \theta $ (see Fig. 1). Thus, the corresponding effective
Hamiltonian is formulated as 
\begin{equation}
E=-\eta N_{T}^{2}(\frac{1}{2}\cos ^{2}\theta +\frac{K}{\eta N_{T}}\sin
\theta \cos \phi +\frac{\gamma }{\eta N_{T}}\cos \theta ).
\end{equation}
Based upon the analysis of a spin in a uniaxial magnetic field \cite
{Chudnovsky3,Zaslavskii,Scharf,Garanin,Garanin1}, we know that there are
stationary states if some angles $(\theta _{0},\phi _{0})$ satisfy $\partial
E/\partial \phi |_{\phi =\phi _{0}}^{\theta =\theta _{0}}=0$ and $\partial
E/\partial \theta |_{\phi =\phi _{0}}^{\theta =\theta _{0}}=0$. The
condition $\partial E/\partial \phi |_{\phi =\phi _{0}}^{\theta =\theta
_{0}}=0$ locates the stationary states in the $XOZ$ plane $(\sin \phi _{0}=0)
$. The existence of multiple stationary states in this quasi-spin system is
equivalent to the existence of multiple metastable MQST states in the
coupled BECs. Near the metastable states the potential describes a
''canyon'' satisfying 
\begin{equation}
E_{\theta }=E(\theta ,\phi _{0})/(\eta N_{T}^{2})=-\frac{1}{2}\cos
^{2}\theta -P\cos (\theta -\theta _{P}).
\end{equation}
The parameters obey $P=\sqrt{K^{2}+\gamma ^{2}}/|\eta N_{T}|$, $\sin \theta
_{P}=K\cos \phi _{0}/\sqrt{K^{2}+\gamma ^{2}}$ and $\cos \theta _{P}=\gamma /%
\sqrt{K^{2}+\gamma ^{2}}$. As stated in the previous section, the parameter $%
K\varpropto \Omega >0$, therefore $\sin \theta _{P}>0$ and $\sin \theta
_{P}<0$ correspond to the equal-phase mode $(\phi _{0}=0)$ and the
anti-phase mode $(\phi _{0}=\pi )$ in the coupled two-component BECs,
respectively. In the case of $E_{2}^{0}-E_{1}^{0}+N_{T}(U_{22}-U_{11})/2=0$,
the parameter $\gamma $ is just the negative detuning $-\delta $, thus $\cos
\theta _{P}>0$ and $\cos \theta _{P}<0$ correspond to the red detuning and
the blue detuning of the coupling laser, respectively. The $\partial
E/\partial \theta |_{\phi =\phi _{0}}^{\theta =\theta _{0}}=0$ is equivalent
to $\partial E_{\theta }/\partial \theta |_{\phi =\phi _{0}}^{\theta =\theta
_{0}}=0$, that is, $\sin 2\theta _{0}+2P\sin (\theta _{0}-\theta _{P})=0$.
For some critical points where both the first and the second derivatives of $%
E_{\theta }$ equal zero, an appreciable tunnelling rate appears. This gives 
\begin{equation}
\begin{array}{l}
\sin 2\theta _{C}+2P_{C}\sin (\theta _{C}-\theta _{P})=0, \\ 
\cos 2\theta _{C}+P_{C}\cos (\theta _{C}-\theta _{P})=0,
\end{array}
\end{equation}
here, $\theta _{C}$ and $P_{C}$ are critical values for $\theta $ and $P$,
respectively. Solving the above equations, one can obtain $\tan ^{3}\theta
_{C}=-\tan \theta _{P}$ and $P_{C}=(\sin ^{2/3}\theta _{P}+\cos ^{2/3}\theta
_{P})^{-3/2}$. The system has an instanton solution at the critical point $%
P=P_{C}$, i.e., $(K/(\eta N_{T}))^{2/3}+(\gamma /(\eta N_{T}))^{2/3}=1$.
This critical point stands on the separatrix between the single stable
regime and the multiple-stable regime. It separates the metastable
multi-MQST behavior between the single-stable population oscillation in the
coupled two-component BEC.

According to the dependence of $E_{\theta }$ on $\theta $, we obtain that
the condition for the existence of multiple stationary states is $P<P_{C}$,
i.e., $(K/(\eta N_{T}))^{2/3}+(\gamma /(\eta N_{T}))^{2/3}<1$. One can
easily find the small oscillations around these stationary states with
nonzero time-averaged values for $S_{z}$ and $\sqrt{S_{x}^{2}+S_{y}^{2}}$.
These oscillations correspond to the phase-locked MQST states with
time-averaged relative phase $0$ or $\pi $ and multiple stationary states
correspond to multiple metastable MQST states with fixed nonzero population
difference and relative phase $0$ or $\pi $. The appearance of multiple
stationary sates indicates, only for some proper parameters, that multiple
metastable MQST states exist. For simplicity we only consider the case where
the parameter $P$ is slightly lower than the fixed critical value $P_{C}$, $%
P=P_{C}(1-\varepsilon )$, $\varepsilon \ll 1$. This requires that the Rabi
frequency, the detuning, the scattering lengths, and the total atomic number
in the coupled BEC system must cooperate with each other to approach the
critical values for the emergence of multiple metastable MQST states. One
way to maintain the critical value $P_{C}$ unchanged is fixing the values of
the ratio $\gamma /K$ and other correlated parameters $(\eta $ and $N_{T})$,
that is, keeping the angle $\theta _{P}$ unchanged. By introducing a new
positive variable $\xi =\theta -\theta _{0}$, the potential $(6)$ can be
expanded into 
\begin{equation}
E_{\theta }(\theta )=E_{\theta }(\theta _{0})+\frac{1}{4}[\sqrt{6\varepsilon 
}\xi ^{2}-\xi ^{3}+O(\xi ^{4})]\sin (2\theta _{C}).
\end{equation}

With the definition in Refs. \cite{Chudnovsky3,Zaslavskii,Garanin,Garanin1},
the tunnelling rate $\Gamma $ obeys $N_{p}(t)=N_{P}(0)\exp (-\Gamma t)$ and
it can be written as $\Gamma =A\exp (-B)$ for the quantum tunnelling regime.
Here, $N_{P}(t)$ is the population occupying the metastable state at time $t$
and the tunnelling exponent $B$ $(\geqq 0)$ is determined by the imaginary
time action of the instanton solution. Similar to Ref. \cite{Chudnovsky3},
the tunnelling exponent follows from the path integral $\int D\{\phi (\tau
)\}\int D\{\cos \theta (\tau )\}\exp (I/\hbar )$ over the continuum of
trajectories which start and end at $(\theta _{0},\phi _{0})$ and which are
close to the instanton solution, where, $\tau $ is the imaginary time $it$,
and $I$ is the imaginary time action $I=\int d\tau [iN_{T}(1-\cos \theta
)d\phi /d\tau +E(\theta ,\phi )]$. Integrating the imaginary time action by
parts, one can gain the tunnelling exponent 
\begin{equation}
\begin{array}{l}
B=N_{T}\int_{-\infty }^{+\infty }d\tau \{\frac{(d\xi /d\tau )^{2}\sin \theta
_{C}}{2P_{C}\sin \theta _{P}}+\frac{1}{4}\sin (2\theta _{C})[\sqrt{%
6\varepsilon }\xi ^{2}-\xi ^{3}+O(\xi ^{4})]\}, \\ 
=16\times 6^{1/4}N_{T}\varepsilon ^{5/4}|\cot \theta _{P}|^{1/6}/5=16\times
6^{1/4}N_{T}\varepsilon ^{5/4}|\gamma /K|^{1/6}/5.
\end{array}
\end{equation}
From the definition of $\varepsilon $, one can obtain 
\begin{equation}
\varepsilon =1-P/P_{C}=1-(1+|\gamma /K|^{2})(1+|\gamma
/K|^{2/3})^{-3/2}|K/(\eta N_{T})|.
\end{equation}
Thus the tunnelling exponent can be expressed as 
\begin{equation}
B=16\times 6^{1/4}N_{T}[1-(1+|\gamma /K|^{2})(1+|\gamma
/K|^{2/3})^{-3/2}|K/(\eta N_{T})|]^{5/4}|\gamma /K|^{1/6}/5.
\end{equation}
To control the tunnelling, one has to select proper values for parameters $%
\gamma $, $K$ and $\eta $. In the experiments performed in a double-well
potential\cite{Orzel,Andrews}, it can be realized by modifying the barrier
position, the barrier height and the magnetic field (using Feshbach
resonances to adjust the scattering lengths, \cite{Inouye}), respectively.
In the experiments with two-component BECs in a single-well potential\cite
{Hall} , it can be realized by adjusting the laser detuning, the laser
intensity, and the magnetic field, respectively. For fixed value of $\eta $
and $\gamma /K$, the tunnelling exponent $B$ decreases with the increasing
of the intensity of the coupling laser. In Fig. 2, we show how the
tunnelling exponent $B$ depends on the angle $\theta _{P}$. In the region
between $0$ and $\pi $, the ratio $B(\theta _{P})/B(\pi /4)$ decreases from
positive infinity to zero when the angle $\theta _{P}$ equals $\pi /2$,
which corresponds to the symmetric case $(\gamma =0)$, and then increases to
positive infinity when the angle $\theta _{P}$ is close to $\pi $. It is
almost flat when the angle $\theta _{P}$ is not close to $0$, $\pi /2$ and $%
\pi $. This angular dependence indicates, in the case of fixed value of $%
\varepsilon $, that the tunnelling exponent increases with increasing $%
|\gamma /K|$.

The result for the angle $\theta _{P}$ close to $\pi /2$, which corresponds
to the symmetric case $\gamma =0$, should be taken with great caution
because the coefficient $\sin (2\theta _{C})$ in the Taylor expansion series 
$(9)$ is equal to zero. In this case, the problem corresponds to the
tunnelling between two equivalent minima which correspond to the angle $%
\theta _{P}$ equal $0$ and $\pi $. Thus the potential can be expanded into
the form of $\xi ^{2}-\xi ^{4}$ and the tunnelling exponent $B$ is expressed
as $B=4S\varepsilon ^{3/2}=4N_{T}\varepsilon ^{3/2}$. Therefore, the
tunnelling exponents $(10)$ and $(12)$ only hold for the asymmetric case
where $\gamma \neq 0$.

To confirm our prediction from the quasi-spin model, we perform a numerical
simulation of the equation (2). A qualitative change in the stationary-state
behavior occurs at $|K/(\eta N_{T})|=1$. When $|K/(\eta N_{T})|>1$, there
are no metastable states for any effective detuning $\gamma $. However, when 
$|K/(\eta N_{T})|<1$, metastable states exist in the region $[-\gamma
_{c},+\gamma _{c}]$ for proper relative phase, where $\gamma _{c}$ satisfies 
$(K/(\eta N_{T}))^{2/3}+(\gamma _{c}/(\eta N_{T}))^{2/3}=1$. See the left
column of Fig. 3. Two stationary states, indicated as $S_{1}$ and $S_{2}$ in
the figure, are stable and the other one $(U)$ is unstable. Adiabatically
changing the effective detuning $\gamma $ from $\gamma _{c}-\varepsilon $ to 
$\gamma _{c}+\varepsilon $ ($\varepsilon $ is a very small positive number),
in the space of the fractional population difference $z=(N_{2}-N_{1})/N_{T}$
and the relative phase $\phi =\alpha _{2}-\alpha _{1}$, a trajectory in the
vicinity of $S_{2}$ becomes a large orbit $C$ encircling $S_{1}$. From the
views of instanton method, the tunnelling exponent is determined by the
canonical action of the orbit, i.e., $B$ follows from the path integral $%
\int D\{z(\tau )\}\int D\{\phi (\tau )\}\exp (I_{c}/\hbar )$ over the
continuum of trajectories which are close to the instanton solution. At
different bifurcation points $\gamma _{c}$, the numerical results show $%
B(|\gamma /K|)/B(|\gamma /K|=1)\varpropto |\gamma /K|^{0.163\pm
0.002}\approx |\gamma /K|^{1/6}$, this confirms our previous prediction from
the quasi-spin model (see the right column of Fig. 3).

There are two important aspects which must be noted. The one is that these
results for tunnelling are only valid in the low barrier limit for the
quasi-spin model, i.e., $\varepsilon <<1$. This means that the above results
only hold in the region which approaches the critical point of emergence of
multiple metastable MQST states. The parametric dependence of the general
case is still an open problem. The other is the validity of the
Wentzel-Kramers-Brillouin (WKB) semiclassical approximation. The
semiclassical approach can only be used in the case of small tunnelling
probability, that is, $B>>1$. In this low barrier limit, from the Taylor
expansion series $(9)$ one can obtain the following tunnelling amplitude by
using the theory developed by Caldeira and Leggett \cite{Caldeira}, 
\begin{equation}
\begin{array}{ll}
A & =(15B/8\pi )^{1/2}\omega , \\ 
& =\eta N_{T}(\frac{15B}{2\pi })^{1/2}(\frac{3\varepsilon }{8})^{1/4}|\cot
\theta _{P}|^{1/6}/(1+\cot ^{2/3}\theta _{P}), \\ 
& =\eta N_{T}(\frac{15B}{2\pi })^{1/2}(\frac{3\varepsilon }{8})^{1/4}|\gamma
/K|^{1/6}/(1+|\gamma /K|^{2/3}).
\end{array}
\end{equation}
Here, $\omega $ is the angular frequency of small oscillations near the
bottom of the inverse potential. Apparently, when the angle $\theta _{P}$ is
close to the $k\pi $ ($k=0,1$), which corresponds to small Rabi frequency or
large detuning of the coupling laser between two BECs, the tunnelling
amplitude $A$ approaches to zero, see Fig. 4. As presented in above, in the
case of $\theta _{P}$ close to $\pi /2$ which corresponds to the symmetric
case $\gamma =0$, the potential is not in form of $\xi ^{2}-\xi ^{3}$ but in
form of $\xi ^{2}-\xi ^{4}$ because the coefficient $\sin (2\theta _{C})$ in
the Taylor expansion series $(9)$ equals zero. Therefore, the above formula
for the tunnelling amplitude only holds for the asymmetric case $\gamma \neq
0$. Generally, contrary to the tunnelling exponent $B$, the tunnelling
amplitude $A$ is sensitive to the structure of quantum levels in the
potential. Therefore, for the case of the full potential $(6)$ and $(7)$,
the estimation of $A$ is still an open problem.

Population transfer between two states in a bistable system can occur either
due to classical thermal activation which depends on the system temperature
or due to quantum tunnelling which does not depend on the system
temperature. There exists a phase transition from the thermal regime to the
quantum regime which occurs at the crossover temperature $T_{C}$. Above $%
T_{C}$, quantum effects are very small and the population transfer rate
follows the Arrhenius law, 
\begin{equation}
\Gamma _{thermal}=\Gamma _{0}\exp (-\frac{U_{B}}{k_{B}T}).
\end{equation}
Here, $U_{B}$ is the height of the energy barrier between two states and $%
k_{B}$ is the Boltzmann constant. Below $T_{C}$, the population transfer is
purely quantum, 
\begin{equation}
\Gamma _{quantum}=A\exp (-B),
\end{equation}
with $B$ independent of the system temperature. Thus the transition occurs
when $\Gamma _{thermal}=\Gamma _{quantum}$. Neglecting the prefactors and
equating the exponents, the crossover temperature can be estimated as 
\begin{equation}
T_{C}=U_{B}/(k_{B}B).
\end{equation}
The transition region is approximately the temperature interval $%
[T_{C}(1-B^{-1}),T_{C}(1+B^{-1})]$. This crossover resembles a first-order
phase transition of the tunnelling rate $\Gamma $ because it is accompanied
with the discontinuity of $d\Gamma /dT$ at $T_{C}$ \cite{Garanin}.

There is another regime for tunnelling, the thermally assisted tunnelling
(TAT), in which the particle strides over the barrier to the bottom of the
potential with lowering temperature \cite{Garanin,Garanin1}. The transition
from the classical regime to the TAT regime resembles a second-order
classical-quantum phase transition of the tunnelling rate $\Gamma $ because
it is accompanied with a discontinuity of $d^{2}\Gamma /dT^{2}$ and no
discontinuity of $d\Gamma /dT$ at the crossover temperature. The
corresponding transition temperature can be estimated as 
\begin{equation}
T_{C}^{/}=\hbar /(\tau _{0}k_{B})=\hslash \omega /(2\pi k_{B}),
\end{equation}
where $\tau _{0}$ and $\omega $ are the period and the angular frequency of
small oscillations near the bottom of the inverse potential, respectively 
\cite{Chudnovsky2,Chudnovsky3,Zaslavskii,Garanin,Garanin1}. In the low
barrier limit $(\varepsilon <<1)$, from the Taylor expansion series $(9)$,
one can obtain the barrier height 
\begin{equation}
U_{B}=\eta N_{T}^{2}(2\varepsilon /3)^{3/2}|\sin (2\theta _{C})|=\frac{5\pi 
}{18}(\hbar /\tau _{0})B,
\end{equation}
where, 
\begin{equation}
|\sin (2\theta _{C})|=2|\gamma /K|_{C}/(1+|\gamma /K|_{C}^{2}).
\end{equation}
Comparing both crossover temperatures, one can easily find that they differ
by a factor $T_{C}/T_{C}^{/}=5\pi /18=1/1.15$, which means that they are of
the same order of magnitude and can both be used to estimate the crossover
temperature.

Below, from the experimental parameters in the experiments of JILA \cite
{Hall}, we will give a quantitative estimation for the tunnelling rate and
the crossover temperature. In those experiments, the atomic mass $%
m_{1}=m_{2}=m_{Rb}=1.45\times 10^{-25}$ kg, the time-averaged orbiting
potential (TOP) magnetic trap has an axial frequency $v_{z}=59$ Hz and a
radial frequency $v_{x,y}=v_{r}=v_{z}/\sqrt{8}=21$ Hz, the $s$-wave scatter
lengths $a_{11}=5.36$ nm, $a_{12}=a_{21}=5.53$ nm and $a_{22}=5.70$ nm, and
the total atomic number $N_{T}\thickapprox 5\times 10^{5}$. To obtain the
numerical values conveniently, we choose the natural units of the problem,
in which, time is in units of $1/(\omega _{x}\omega _{y}\omega _{z})^{1/3}=1/%
\overline{\omega }$, length is in units of the size of the
geometric-averaged harmonic-oscillator length $\overline{d}=\sqrt{\hbar
/[(\omega _{x}\omega _{y}\omega _{z})^{1/3}m_{Rb}]}=\sqrt{\hbar /(\overline{%
\omega }m_{Rb})}$, energy is in units of the geometric-averaged trap level
spacing $\hbar (\omega _{x}\omega _{y}\omega _{z})^{1/3}=\hbar \overline{%
\omega }$, and mass is in units of Rb atomic mass $m_{Rb}$.

Due to gravity acting besides the TOP, the centers of two condensates will
displace along the vertical direction and the two equilibrium displacements
are generally not the same. Thus, if the interparticle interaction is
absent, the lowest single-particle state has the familiar wave function, 
\begin{equation}
\Phi _{0i}(\stackrel{\rightharpoonup }{r})=\frac{1}{\pi
^{3/4}(d_{x}d_{y}d_{z})^{1/2}}\exp (-\frac{x^{2}}{2d_{x}^{2}}-\frac{y^{2}}{%
2d_{y}^{2}}-\frac{(z-\digamma _{i}z_{0})^{2}}{2d_{z}^{2}}).
\end{equation}
Where, $\digamma _{1}=+1$, $\digamma _{2}=-1$, $2z_{0}$ is the offset
between two potential centers along the vertical axis, $d_{k}=\sqrt{\hbar
/(\omega _{k}m_{Rb})}$ $(k=x,y,z)$ are the oscillator lengths. The offset $%
2z_{0}$ between two condensates can be varied by adjusting the magnitude of
the rotating magnetic field. In the presence of interatomic interaction, the
dimensions of the condensates are changed. The spatial parts of the
macroscopic quantum wave functions are in the shape of 
\begin{equation}
\Phi _{i}(\stackrel{\rightharpoonup }{r})=\frac{1}{\pi
^{3/4}(b_{ix}b_{iy}b_{iz})^{1/2}}\exp (-\frac{x^{2}}{2b_{ix}^{2}}-\frac{y^{2}%
}{2b_{iy}^{2}}-\frac{(z-\digamma _{i}z_{0})^{2}}{2b_{iz}^{2}}).
\end{equation}
The variational parameters $b_{ik}$ $(k=x,y,z;$ $i=1,2)$ depend on the
scattering length, the total atom number, and the trapping potential and
they have almost the same numerical values as $d_{k}$. For proper values of
the offset $2z_{0}$, the numerical results of \cite{Williams} show that the
spatial distributions $\Phi _{i}(\stackrel{\rightharpoonup }{r})$ and their
overlap only weakly depend on the total atom numbers in each condensate. For
simplicity, in the following calculations, the variational parameters $%
b_{ik} $ are replaced by the oscillator lengths $d_{k}$. Therefore, the
parameters $E_{i}^{0}$, $U_{ij}$ and $K$ are determined by 
\begin{equation}
\begin{array}{ll}
E_{1}^{0} & =E_{2}^{0}=\hbar (\omega _{x}+\omega _{y}+\omega _{z})/2, \\ 
U_{ii} & =4\pi \hbar ^{2}a_{ii}/[(\sqrt{2\pi }%
)^{3}d_{x}d_{y}d_{z}m_{Rb}],(i=1,2), \\ 
U_{12} & =4\pi \hbar ^{2}a_{12}\exp (-2z_{0}^{2}/d_{z}^{2})/[(\sqrt{2\pi }%
)^{3}d_{x}d_{y}d_{z}m_{Rb}], \\ 
K & =\hbar \Omega \exp (-z_{0}^{2}/d_{z}^{2}).
\end{array}
\end{equation}
So the corresponding parameters in the quasi-spin model $(5)$ can be written
as $\gamma =\hbar ^{2}N_{T}(a_{22}-a_{11})/(\sqrt{2\pi }%
d_{x}d_{y}d_{z}m_{Rb})-\hbar \delta $ and $\eta =\hbar
^{2}[a_{22}+a_{11}-2a_{12}\exp (-2z_{0}^{2}/d_{z}^{2})]/(\sqrt{2\pi }%
d_{x}d_{y}d_{z}m_{Rb})$. In the case of complete overlap $(2z_{0}=0)$, the
anisotropy parameter $\eta $ equals zero, thus the metastable multi-MQST
behavior will never appear, but some running-phase MQST states may still
exist. This indicates that, to insure the existence of multiple metastable
MQST states, a finite offset must be kept between two condensates.
Furthermore, the appearance of this kind of MQST requires $K^{2/3}+\gamma
^{2/3}<(\eta N_{T})^{2/3}$. Because $K\varpropto \Omega $ and $\gamma
\varpropto \delta $, this inequality indicates that the Rabi frequency and
the detuning of the coupling pulses must be relatively small. Choosing the
total atom number $N_{T}=2.0\times 10^{4}$, the half offset $z_{0}=0.20d_{z}$%
, the Rabi frequency $\Omega =2\pi \times 10$ Hz, and the detuning $\delta
=-179$ Hz, one can get $\eta N_{T}=6.70\times 10^{-32}$, $\gamma =4.57\times
10^{-32}$, $K=6.90\times 10^{-33}$ and $\varepsilon =9.78\times 10^{-3}$.
Thus, the corresponding tunnelling exponent $B$ and crossover temperature $%
T_{C}$ are around $4.22\times 10^{2}$ and $3.54\times 10^{-2}$ nK,
respectively. Obviously, the crossover temperature $T_{C}$, which
corresponds to a phase transition from classical tunnelling to quantum
tunnelling, is far below than the critical temperature $T_{0}\thickapprox
150 $ nK for Bose-Einstein condensation in a dilute gas of $^{87}Rb$.

\section{Discussion and summary}

The generalized Bloch equation $(4)$ and its stability analysis will help to
control the population transfer and realize the single-qubit operation with
BECs qubit. Theoretically, any two-state quantum system can serve as a
qubit, many of them have been realized experimentally. To make use of two
quantum states, the coherence and superposition between them is the most
essential qualification. The experimental observation of coherence and
superposition between two BECs indicates the possibility of encoding two
coupled BECs as a qubit. However, because of the mean-field interaction
among Bosonic condensed atoms, the qubit operations become very difficult to
perform. To accomplish a single-qubit operation, it must be possible rotated
arbitrarily in the Hilbert space. This requires the atomic populations can
be transferred arbitrarily. From the Bloch equations (4), we find that MQST
prevents the arbitrary rotation of the state vector. And even if there no
MQST, when $\eta \neq 0$, the complete population inversion can not be
accomplished with linear operations. Thus, to accomplish a linear qubit
operation, one has to adjust the parameter $\eta $ to zero by varying the
atomic scattering length with a Feshbach resonance \cite{Inouye}. In this
case, the mean-field interaction gives a density-shift to the original
energy levels and, according to Rabi's theory, the arbitrary rotation of the
state vector can be performed easily. Thus, if one encodes the qubit states $%
\left| 0\right\rangle $ and $\left| 1\right\rangle $ as the condensate
wavefunctions for two condensates in a double-well potential or two
hyperfine-state condensates coupled with Raman pulses \cite{Shi}, an
arbitrary one-bit linear operation can be realized when the anisotropy is
absent $(\eta =0)$ and an arbitrary one-bit nonlinear operation can be
realized when the metastable multi-MQST behavior is absent $(|K|>|\eta
N_{T}|)$. This means that, to perform an arbitrary one-bit transformation,
it at least needs choosing proper parameters to avoid the emergence of the
metastable multi-MQST behavior.

The tunnelling of the quasi-spin model described by the Hamiltonian $(5)$
has also been investigated by mapping it onto a particle moving in an
asymmetric double-well potential \cite{Zaslavskii,Scharf,Garanin}. Using
this approach, Garanin et. al. have explored some new fascinating feature of
this uniaxial spin model in the strongly biased limit \cite{Garanin}. They
find that there exist two different regimes for the classical-quantum
transition of the tunnelling rate and the kind of transition depends on both
the strength and the direction of the magnetic field. In this article, we
directly analyze the tunnelling in the low barrier limit for the quasi-spin
model, which corresponds to the effective magnetic fields near their
critical values for appearance of metastable states. This requires that all
physical parameters of the coupled BECs collaborate with each other to
approach the critical point of appearance of multiple metastable MQST
states. The symmetric case $(\gamma =0)$ of the coupled BECs corresponds to
the unbiased case $(H_{z}=0)$ of the anisotropic spin model, which has been
investigated in details by mapping it onto a particle moving in a symmetric
double-well potential \cite{Chudnovsky2}.

The macroscopic quantum tunnelling of two-component BECs has also been
investigated by the Kasamatsu group. Using a numerical approach, they have
analyzed the tunnelling between two kinds of metastable stationary states, a
symmetry-breaking state (SBS) and a symmetry-preserving state (SPS), in
uncoupled two-component BECs \cite{Kasamatsu}. To improve the usual Gaussian
variational method, they have introduced a collective coordinate approach
and then calculated the tunnelling rate within the WKB approximation. In
that system, the populations of the two components can not be converted into
each other because of the absence of coupling. This means, the tunneling
does not occur between two components but between stationary states with
different spatial configurations. Thus, this kind of tunnelling originates
from the quantized spatial structure of the Hamiltonian. In our model, due
to the coupling, the population can be transferred from one component to the
other. Furthermore, we assume the coupling is very weak, thus both
components stay in their ground stationary states through the full process.
The metastability (metastable MQST) is the result of the cooperation between
the coupling and the mean-field interaction (including both the
intra-component and the inter-component interaction). Correspondingly, the
tunneling from the metastable self-trapped state to its ground state of the
coupled two-component BECs is caused by the quantized structure of their
field operators.

In conclusion a system of coupled BECs (two BECs in a double-well potential
or two internal state BECs coupled with laser pulses) has been mapped to a
spin in a magnetic field by introducing a generalized Bloch vector. The
mean-field interaction, the coupling and the asymmetry or the detuning are
relevant to the anisotropy, the transverse magnetic field and the
longitudinal magnetic field, respectively. The corresponding generalized
Bloch equation is obtained. The analysis of this generalized Bloch equation
will be propitious to control the population transfer and realize the
quantum computation with coupled BECs. Based upon experience from the
well-studied tunneling of spin systems, the detailed information about the
tunnelling between two metastable MQST states in coupled two-component BECs
can be obtained with the imaginary-time path-integral method. The crossover
temperature $T_{C}$ at the critical point for a transition from the
classical thermal regime to the quantum regime was obtained. When the system
temperature decreases through $T_{C}$, the population conversion goes from
classical thermal activation regime to purely quantum tunnelling regime.
This means, below the crossover temperature $T_{C}$, the quantum
fluctuations in the atomic fields take the dominant position. We also find
that the tunnelling rate can be adjusted by varying the coupling and the
trapping magnetic field.

\begin{center}
{\Large Acknowledgment}
\end{center}

The author Lee is very grateful to the help of Prof. X. Zhu in WIPM and the
comments from Dr. J. Brand in MPI-PKS. The work is supported by NSFC (Grant
No. 10275023 and 10274093), National Fundamental Research Program (Grant No.
2001CB309300) and foundations of CAS and MPI-PKS.

\begin{center}
{\Large Figure capt}{\large ion}
\end{center}

\begin{quotation}
Fig. 1 The quai-spin $\stackrel{\rightharpoonup }{S}$ and it's components $%
(u,v,w)$ in conventional spherical coordinates.

Fig. 2 The tunnelling exponent ratio $B(\theta _{P})/B(\theta _{P}=\pi /4)$
versus different $\theta _{P}$. Where, the angle $\theta _{P}$ characterizes
the angle between the effective magnetic field $\stackrel{\rightharpoonup }{B%
}_{eff}=K\stackrel{\rightharpoonup }{e}_{x}+\gamma \stackrel{\rightharpoonup 
}{e}_{z}$ and axis-$z$.

Fig. 3 In the left column, the stationary states for $|K|<|\eta N_{T}|$ are
shown. There are two metastable states $S_{1}$, $S_{2}$ and one unstable
state $U$. In the right column, the tunnelling exponent ratio $B/B_{0}$
versus different $|\gamma /K|$ is presented, where $B_{0}=B(|\gamma /K|=1)$.
The black dots show the numerical data and the straight line represents the
linear fit for the logarithmic data.

Fig. 4 The tunnelling amplitude ratio $A(\theta _{P})/A(\theta _{P}=\pi /4)$
versus different $\theta _{P}$, where the angle $\theta _{P}$ characterizes
the angle between the effective magnetic field $\stackrel{\rightharpoonup }{B%
}_{eff}=K\stackrel{\rightharpoonup }{e}_{x}+\gamma \stackrel{\rightharpoonup 
}{e}_{z}$ and axis-$z$.
\end{quotation}

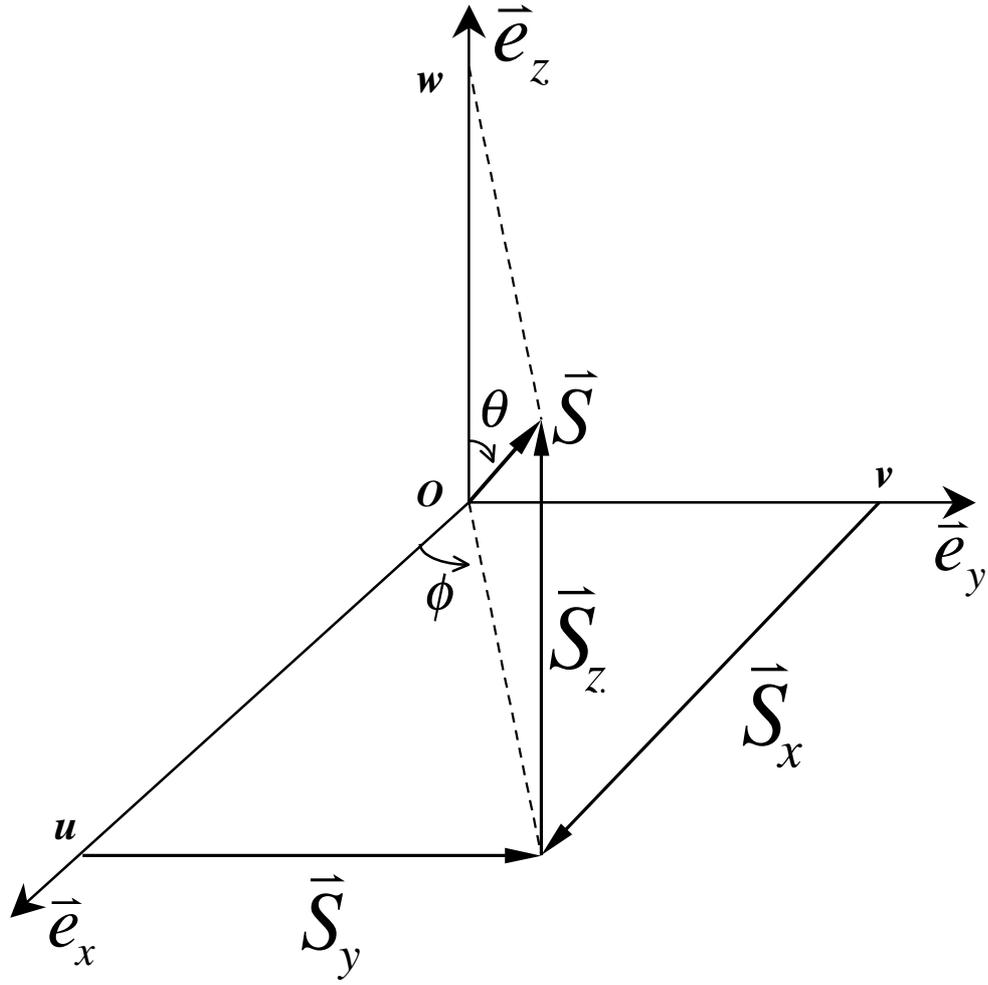
\end{document}